\documentclass[a4paper,11pt]{article}
\pdfoutput=1 
\usepackage{jinstpub}

\usepackage[utf8]{inputenc}
\usepackage[english]{babel}

\usepackage{graphicx}
\usepackage{csquotes}

\title{Electron evaporation from magnetic trap in Troitsk nu-mass experiment}
\author[1,2]{Alexander Nozik}
\author[1,2]{Timur Hamitov}

\affiliation[1]{Moscow Institute of Physics and Technology, Institutsky lane 9, Dolgoprudny, Moscow region, 141700}
\affiliation[2]{Institute for Nuclear Research of the Russian Academy of Sciences (INR RAS), Prospekt 60-letiya Oktyabrya 7a, Moscow, Russia, 117312}

\emailAdd{nozik.aa@mipt.ru}

\abstract{
    This paper is dedicated to the simulation of the so-called trapping-effect observed in the Troitsk nu-mass experiment. The effect is caused by the magnetic trapping of decay electrons in the windowless gaseous tritium source and the gradual evaporation of those electrons. As a result, alongside regular tritium beta-spectrum electrons, we see additional electrons that are initially trapped in the source and escape it with changed energy. The spectrum of evaporated electrons is quite peculiar (almost flat for monochromatic initial electrons) and could not be directly measured in the experiment. So one has to rely on simulations. Also, it is possible that the same effect could be observed in other cases of magnetic traps.
}

\begin{document}

\maketitle

\section{Introduction}

\begin{figure}
    \centering
    \includegraphics[width = 0.95\linewidth]{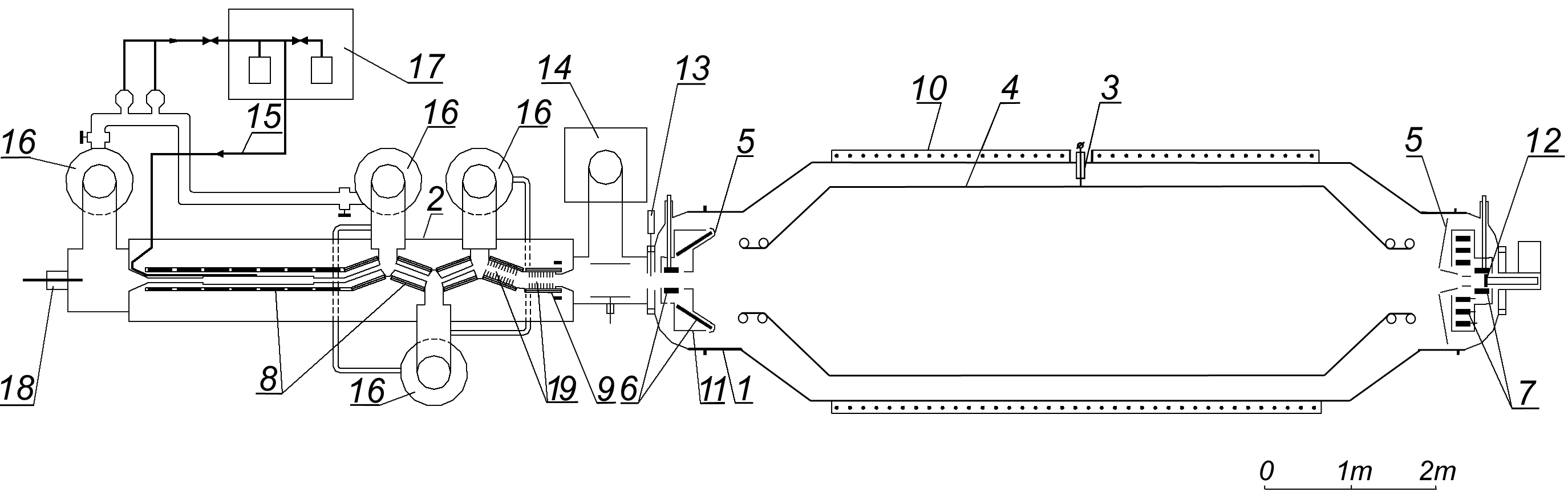}
    \caption{Troitsk nu-mass setup. The tritium source is marked by number 8. The spectrometer electrode is marked by number 4. Pinch-magnet is marked by number 6.}
    \label{fig:setup}
\end{figure}

The Troitsk nu-mass experiment is dedicated to the search for sterile neutrinos. The setup consists of a windowless gaseous tritium source (WGTS), electrostatic spectrometer with adiabatic magnetic collimation (MAC-E spectrometer), and a semiconductor detector (Fig.~\ref{fig:setup}).The setup with a smaller spectrometer was used in measurements of the mass of electron anti-neutrino before 2010 (\cite{Aseev:2011dq, Belesev:2012hx, Belesev:2013cba, Nozik:2019jgm}) and the larger version was used after 2010 to search for sterile neutrinos (\cite{Abdurashitov:2015jha, Abdurashitov:2017kka}). The tritium source remained the same.

The principle of operation is the following:
\begin{enumerate}
    \item Electrons are produced in a beta-decay in a WGTS decay volume.
    \item Then they are transported via an adiabatic magnetic transport system to the spectrometer. During transport inside the decay volume they could scatter on the tritium molecules (the pressure of the gas is low, but the scattering probability on one pass still could rise to $\sim 0.5 $).
    \item In the so-called pinch-magnet at the entrance of the spectrometer (nominal field $7.2~T$) the electrons have the maximum angle between their velocity and the spectrometer axis.
    \item The field in the center of the spectrometer (so-called analyzer plane) is down to $10^{-3}~T$. Due to conservation of adiabatic invariant the transversal component of electron velocities in the analyzer plane is quite small (up to $10^{-4}$ of its full energy).
    \item The stopping electric field is applied near the analyzing plane parallel to the spectrometer axis. All electrons with energy higher than the stopping field power pass the spectrometer and are registered by the detector. All others are reflected. The spectrum measurements are done by making scans with the stopping field.
\end{enumerate}

The term "trapping effect" was introduced and first discussed in \cite{Lobashev:1999tp}. The magnetic trap is a region in space with the magnetic field limited from both sides with a higher magnetic field. A higher magnetic field causes so-called magnetic reflection for electrons with specific angles between their velocities and the trap axis.  There are two large traps in the setup: the spectrometer itself and the WGTS decay volume. 

The spectrometer is supposed to be "clean" meaning that it does not contain radioactive materials that could produce high-energy electrons. Low-energy electrons are seldom trapped in it, but they do not affect measurements as long as their energy is lower than the detection threshold. Occasional trapped high energy electrons from cosmic rays produce so-called "bunches" - short signals of higher count rate. Those bunches are treated during the signal analysis and do not critically affect the resulting spectrum (see \cite{Lobashev:1999tp} and \cite{Aseev:2011dq} for details).


The trap in the source on the other hand is constantly fed from tritium decays with large angles. The magnetic bottle scheme showed in Fig.~\ref{fig:trap}. Electrons with large angles relative to the bottle axis are trapped indefinitely unless they suffer scattering on the tritium gas and their angle is changed. The tritium could not be removed from the source since it is used to produce the electrons in the first place. So those electrons keep escaping from the trap with the constant rate, thus creating an irremovable physical background with a distorted original spectrum.  

In this article, we will discuss trapping-effect physics and simulation results. 

\section{Trapping physics}

\begin{figure}
    \centering
    \includegraphics[width = 0.9\textwidth]{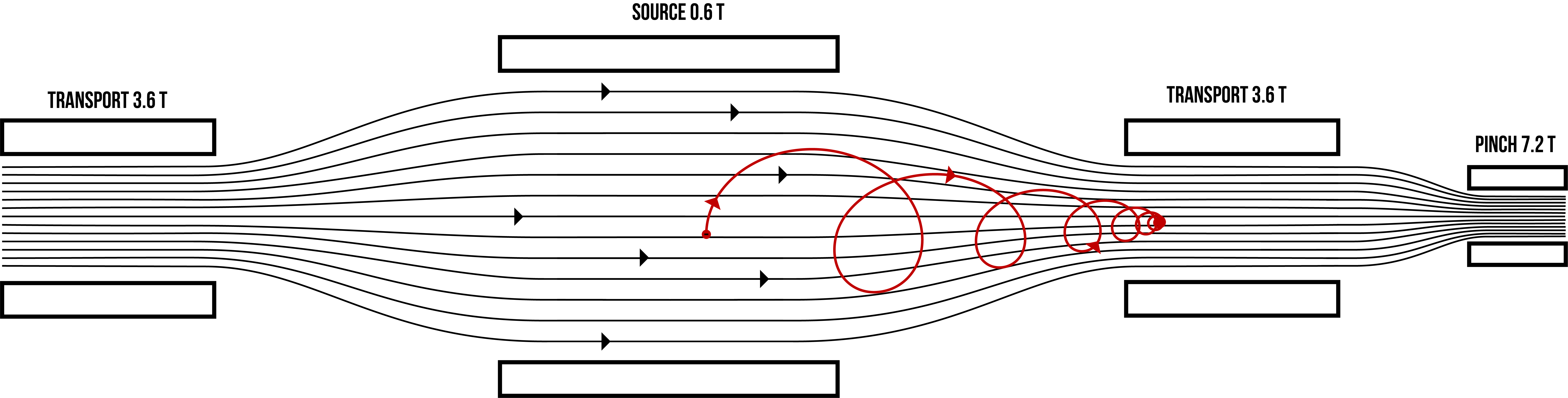}
    \caption{Troitsk nu-mass tritium source magnetic field}
    \label{fig:trap}
\end{figure}

Electrons are produced during $\beta$-decay of tritium in a so-called tritium source which is a tube 3 m long and 5 cm in diameter. The source has $0.6~T$ axial field value, and there is a stronger field (with a nominal value of $3.6~T$), created by transport magnets. Those fields are used to adiabatically transport electrons to the spectrometer (field scheme is shown at Fig.~\ref{fig:trap}). Adiabatic mode means the conservation of invariant $\mu = \frac{-mv_ \perp ^ 2} {2B}=const $. Here $v_ \perp$ is the velocity component, perpendicular to the field $B$, $m$ is the electron mass. 

\begin{figure}
    \centering
    \includegraphics[width = 0.5\textwidth]{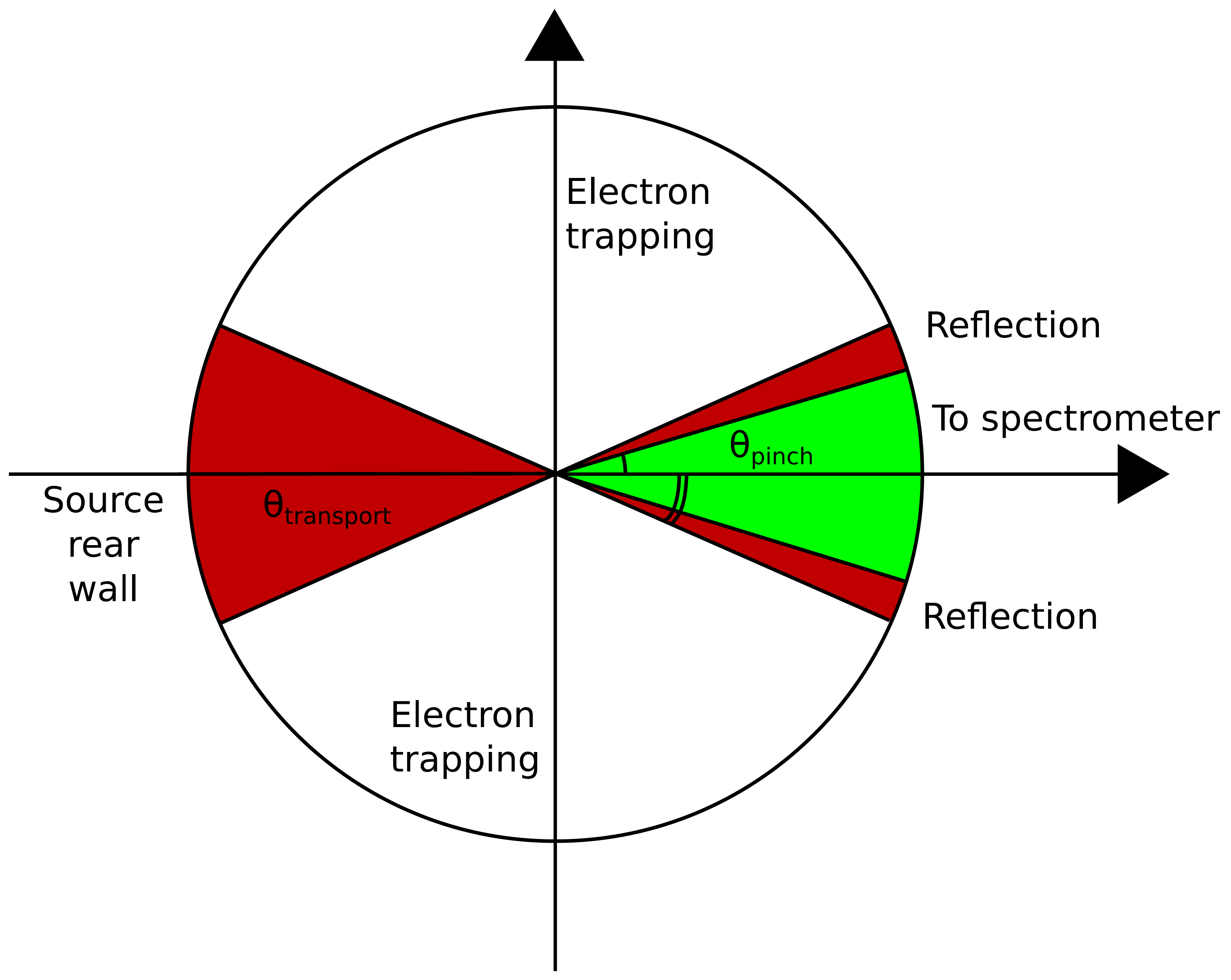}
    \caption{The principal scheme for trapping. $\theta$ is the angle between the magnetic field axis and electron's velocity. }
    \label{fig:scheme}
\end{figure}

The principal scheme of the trapping is shown as an angle diagram at Fig.~\ref{fig:scheme}. In this scheme, we show only angles between the electron velocity direction and the trap axis. To simplify the explanation we assume that the density of the gas in the trap is small (the pressure in the source is up to $10^{-4}~mbar$), so if after being born or after scattering electron has some angle relative to the trap axis, it will have the same angle near the trap boundary. Later, the simulation takes into account the density effect, but it does not change the picture dramatically. Electrons are supposed to be born isotropically. The electrons in the left red zone leave the trap at the rear side and are lost. Electrons in the right green region are accepted into the spectrometer. Electrons in the top and bottom white regions are trapped in the magnetic bottle. The red gap to the right is specific to Troitsk nu-mass: electron could escape the trap in the tritium source, but be reflected from pinch magnet. In this case, they pass back through the source and escape through the rear trap exit and are lost.

The reflection angles are determined by magnetic field ratio:
\begin{gather}
    \theta_{pinch} = \arcsin \sqrt{\frac{B_{source}}{B_{pinch}}} = 16.8^\circ,\\
    \theta_{transport} = \arcsin \sqrt{\frac{B_{source}}{B_{transport}}} = 24.1^\circ,
\end{gather}
where $B_{source}$ is the axial field in the tritium source (trap body), $B_{transport}$ - field in transport channels and $B_{pinch}$ - field inside the pinch magnet. Thus, we get that only about 5\% of electrons are accepted by the spectrometer and more than 80\% of all produced electrons are trapped. The remaining 15\% escapes through the rear magnetic mirror. The direct simulation of escaped electrons show that the chance of escaped electrons to get back into the source is rather small (less than $10^{-5}$), so they could be ignored at this level of precision. The number could be different for different field configurations. For example in KATRIN experiment (\cite{Schonung:2016tal}), where is no magnetic reflection, but the back-scattering from the rear wall is the dominant effect.

The electrons scatter on the residual gas and change its angle and energy. They can escape the trap in two ways:
\begin{itemize}
    \item Drift to the right side of Fig.~\ref{fig:scheme}, then jump the gap in one single interaction and arrive in the green accepted zone (pass to the spectrometer).
    \item Drift to the left and escape through the rear mirror.
\end{itemize}

The electrons that go to the right green zone, but could not cross the gap in one interaction are reflected from the pinch magnet. In most cases, those electrons pass through the source and automatically fall through the rear mirror (since the chance of interaction in one pass is small). 
The interaction of electrons with residual gas is described by three processes:
\begin{itemize}
    \item Quasi-elastic scattering on a nucleus. In this process, the energy change is rather small (fractions of an eV). At the same time, Coulomb scattering could produce significant angles (more than $3.65^\circ$ required to jump the gap).
    \item Ionization of the hydrogen molecule. In this case, the angle change is minimal, but the energy loss probability is inversely proportional to the loss value squared and could rise to 100 eV (more details in \cite{Abdurashitov:2016nrv}).
    \item Excitation of the hydrogen molecule. Works the same way as ionization, but with energy loss equal to the energy difference between levels (up to 15.4 eV).
\end{itemize}

The excitation and ionization losses behave similarly, so they could be joined in a group called inelastic scattering. While the elastic cross-section shows the rate of angle change, the ratio between elastic and inelastic cross-sections shows the amount of "friction", the loss of energy per one angle-changing scattering.

The electron drifts in angular space towards the escape angle (both forward and backward) and then goes out with smaller energy. This process could be roughly described as evaporation in angular space.

\section{Simulation}

\begin{figure}
    \centering
    \includegraphics[width = 0.9\textwidth]{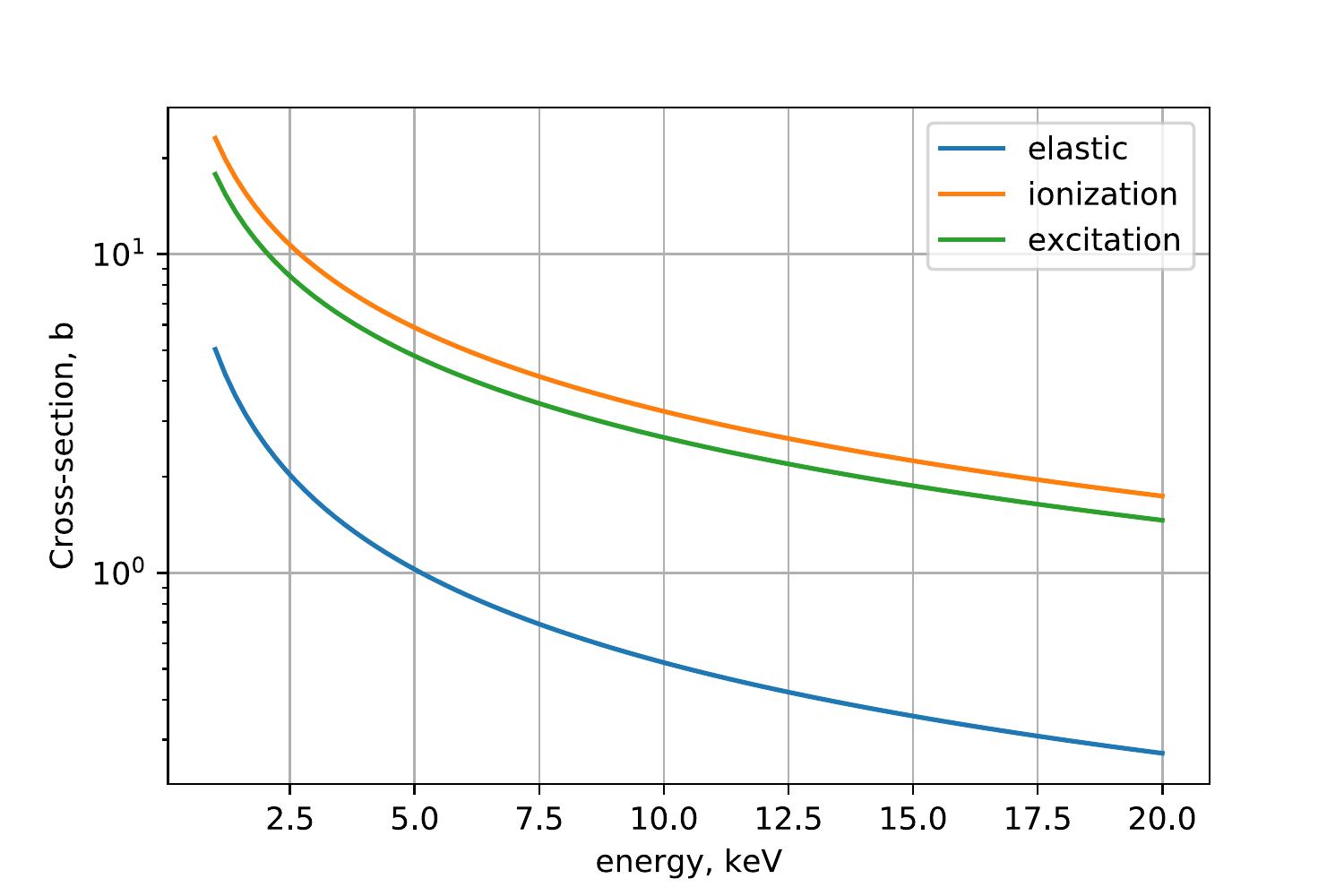}
    \caption{Quasi-elastic, ionization and excitation cross-section dependency on incident electron energy.}
    \label{fig:cross-sections}
\end{figure}

The simulation was implemented with the code written in Kotlin language (\cite{Nozik:2019gmi, Nozik:2020wiy}) and utilizes heavy-duty parallel computations for electron propagation. The simulation uses a simplified two-dimensional phase-space geometry tracking only electron energy ($E$), the angle between electron velocity and trap axis ($\theta$), and position along trap axis ($z$). The distribution of electrons across the plane perpendicular to the trap axis is assumed to be uniform. Non-uniformity of the magnetic field near the traps causes a small transversal drift of electrons. But the quantitative estimation shows that the time required for this drift to be significant is much larger than the time needed for an electron to escape the trap or for its energy to drop below the acceptable energy. A polar angle is taken into account during scattering but assumed to be random (it changes continuously during the motion due to rotation around a magnetic field line). 

The code for scattering cross-sections, angle change, and energy loss computation was provided by Ferenc Gluck and Sebastian Voecking (\cite{kasiopea}). Computed cross-sections are shown at \ref{fig:cross-sections}. It was rewritten in Kotlin and optimized for parallel computations (\cite{trapping-repo}).

The simulation has following input parameters:
\begin{itemize}
    \item Initial electron energy.
    \item Low electron energy cutoff. When electron energy drops below this value, the simulation stops. The parameter is introduced to reduce computation time.
    \item Maximum magnetic fields in the source, transport magnets and pinch-magnet: $B_{source}$, $B_{transport}$ and $B_{pinch}$. 
    \item Gas density to calculate the free path.
    \item Optionally, a full field map could be provided as well. If it is provided, the angles after scattering are calculated using a field in a specific scattering point (the transport between scattering points is still adiabatic). If the field map is not provided, the field is considered to be uniform. The introduction of the field map reduces the computation speed but does not significantly affect the results.
\end{itemize}

At the end of the simulation, an electron could have one of the following statuses:
\begin{itemize}
    \item \textbf{PASS} - the electron has the angle accepted by the spectrometer without scattering.
    \item \textbf{REJECTED} - the electron leaves the trap through the rear plug with or without scattering.
    \item \textbf{ACCEPTED} - electrons in spectrometer acceptance angle after at least one scattering.
    \item \textbf{LOWENERGY} - electrons with energy under the energy cut.
\end{itemize}

\begin{figure}
    \centering
    \includegraphics[width = 0.9\textwidth]{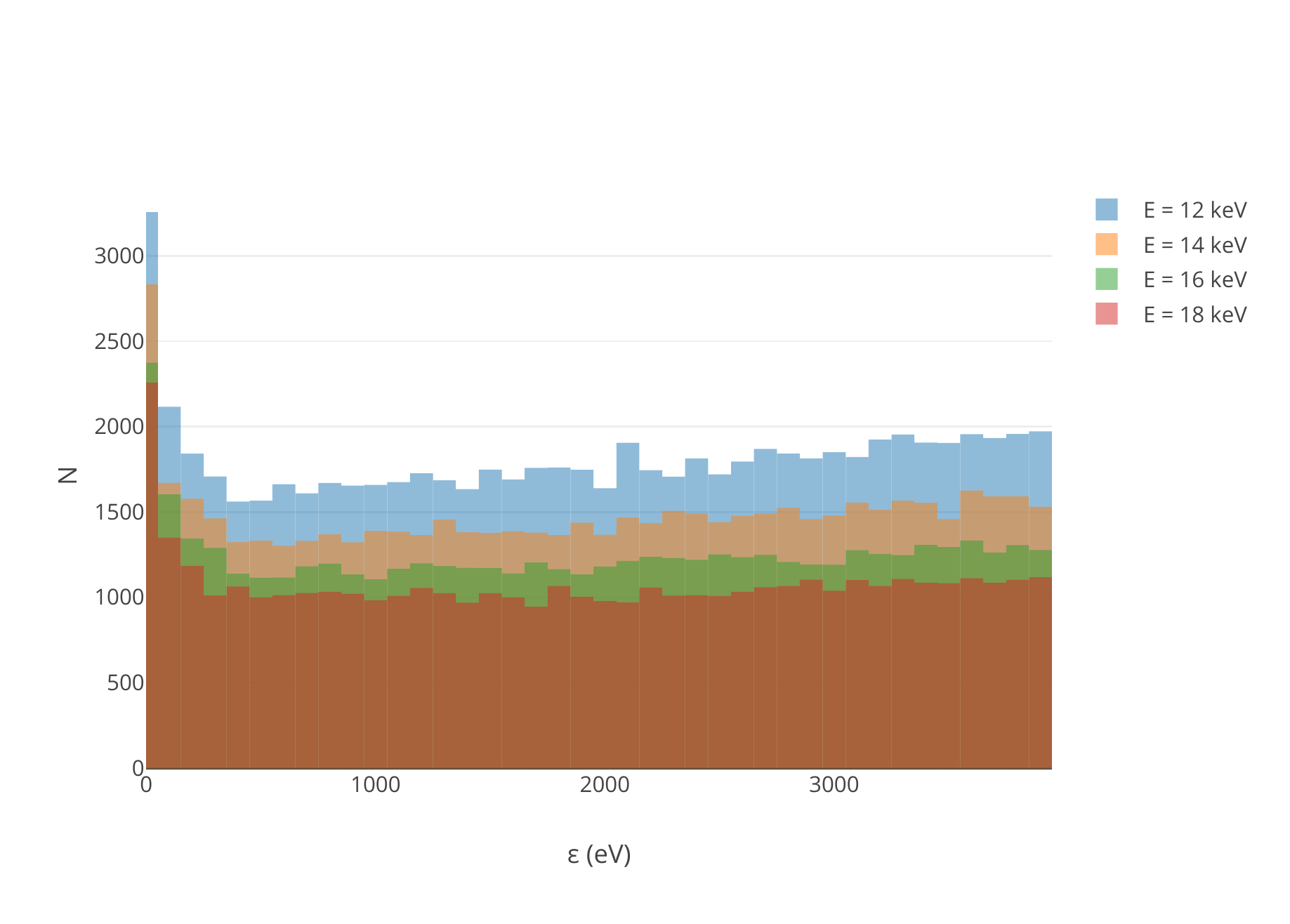}
    \caption{The energy spectra of escaped electrons for different starting energy. $\varepsilon = E_{initial} - E_{escaped}$ is the loss of energy. Vertical axis shows the number of electrons per 100 eV energy bin for $10^7$ initial electrons.}
    \label{fig:energy-spectra}
\end{figure}

\begin{figure}
    \centering
    \includegraphics[width = 0.7\textwidth]{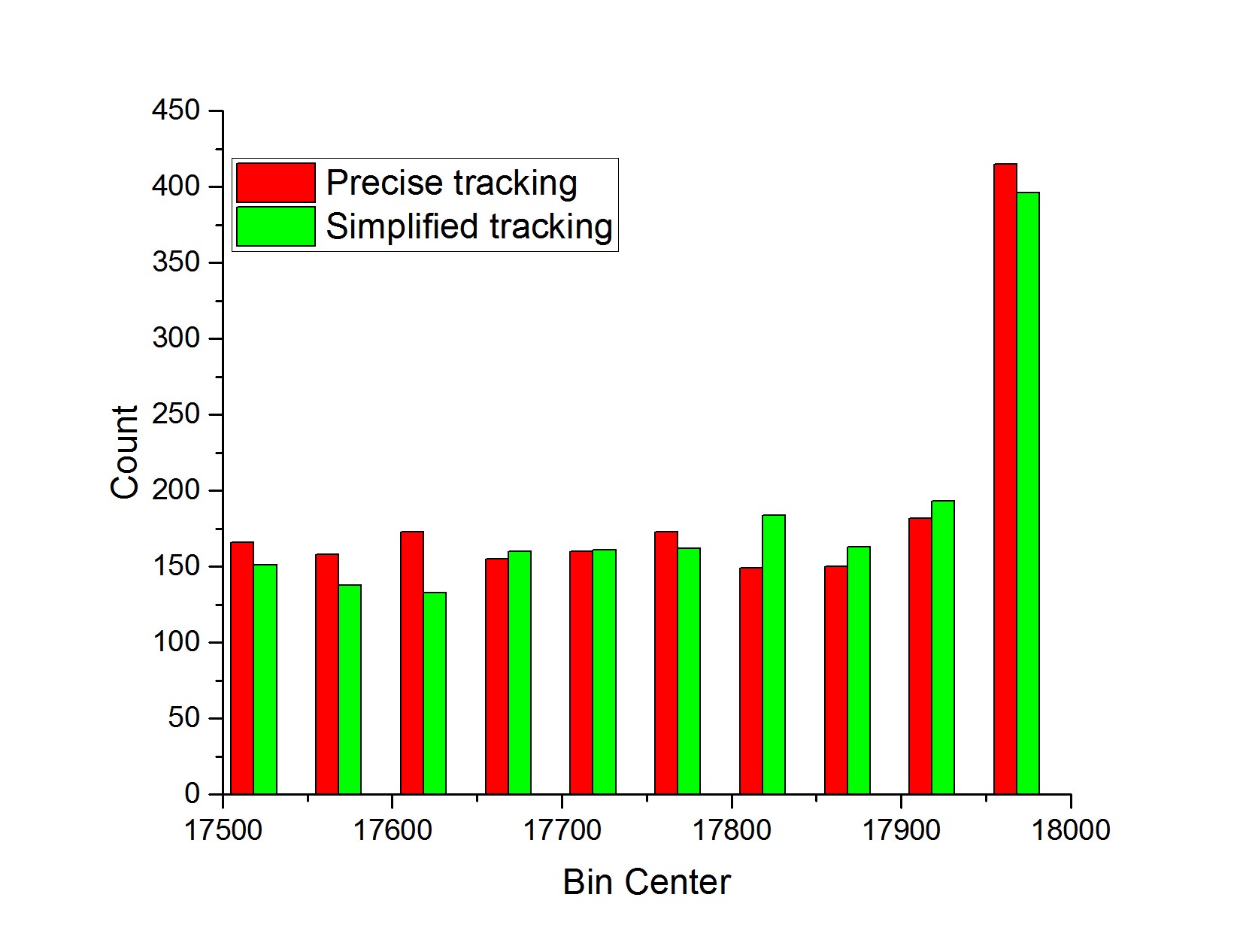}
    \caption{Comparison of full-tracking simulation with simplified model. The normalized number of escaped electrons in 50 eV energy bin versus electron energy in eV.}
    \label{fig:track-compare}
\end{figure}

The Fig.~\ref{fig:energy-spectra} shows the resulting energy distributions for \textbf{ACCEPTED} electrons with fixed starting energies from 12 to 18 keV. In each case, we take mono-energetic starting distribution. It could be later convoluted with actual energy distribution if needed. The spectrum is characterized by a small rise near zero losses which is caused by electrons that suffer only one or several scattering before hitting the acceptance zone. This part of the spectrum mostly follows $1/\varepsilon^2$ law of ionization loss cross-section. The rest of the region is represented by an almost flat curve (the height depends only on initial electron energy). Due to this flatness, it was possible to replace the trapping effect with a simplified formula (with a constant rate for each energy) at \cite{Aseev:2011dq} and \cite{Abdurashitov:2017kka}.

The simulation utilized the simplified geometric model and supposes that reflection from the magnetic bottleneck is instantaneous. In reality, the magnetic field rises from $B_{source}$ to $B_{transport}$ gradually on the length of approximately 10 cm. The scattering in magnetic plug produces electrons with different angular distribution so this process could in theory affect the resulting energy distribution of escaped electrons. To check this possibility and take into account the three-dimensional picture, a single simulation was performed with full electromagnetic three-dimensional tracking. The energy range for this simulation was significantly reduced to allow feasible simulation time. The precise tracking simulation was performed by Aino Skasyrskaya at INR RAS. The Fig.~\ref{fig:track-compare} shows the result of this calculation. It could be seen that results of precise and simplified tracking match within statistical errors. 

\section{Discussion}

\begin{figure}
    \begin{minipage}{0.45\linewidth}
        \centering
        \includegraphics[width=\linewidth]{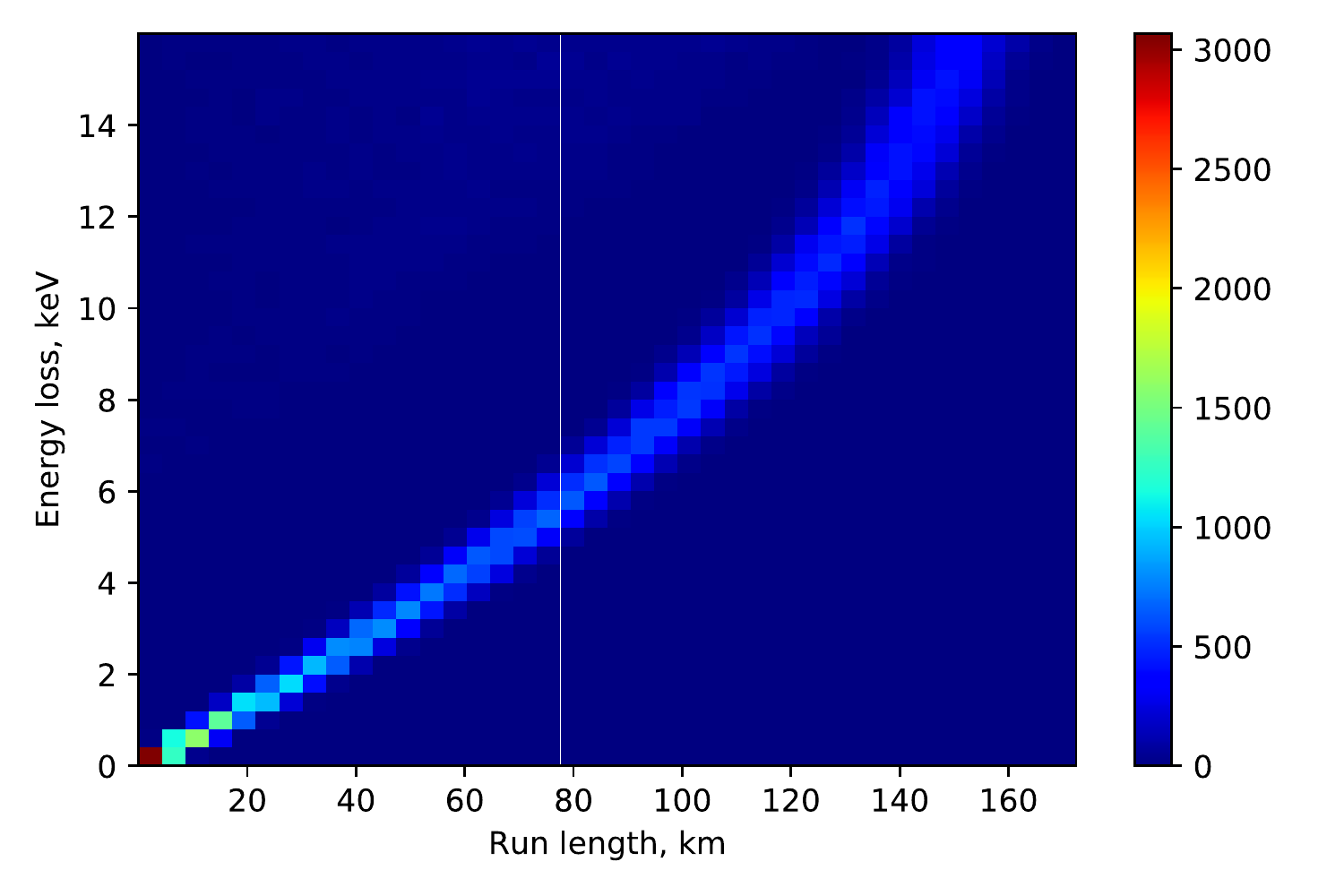}
        \caption{Energy loss versus full electron run length for ACCEPTED electrons. The initial energy was 20 keV. }
        \label{fig:loss-vs-length}
    \end{minipage}
    ~
    \begin{minipage}{0.45\linewidth}
        \centering
        \includegraphics[width=\linewidth]{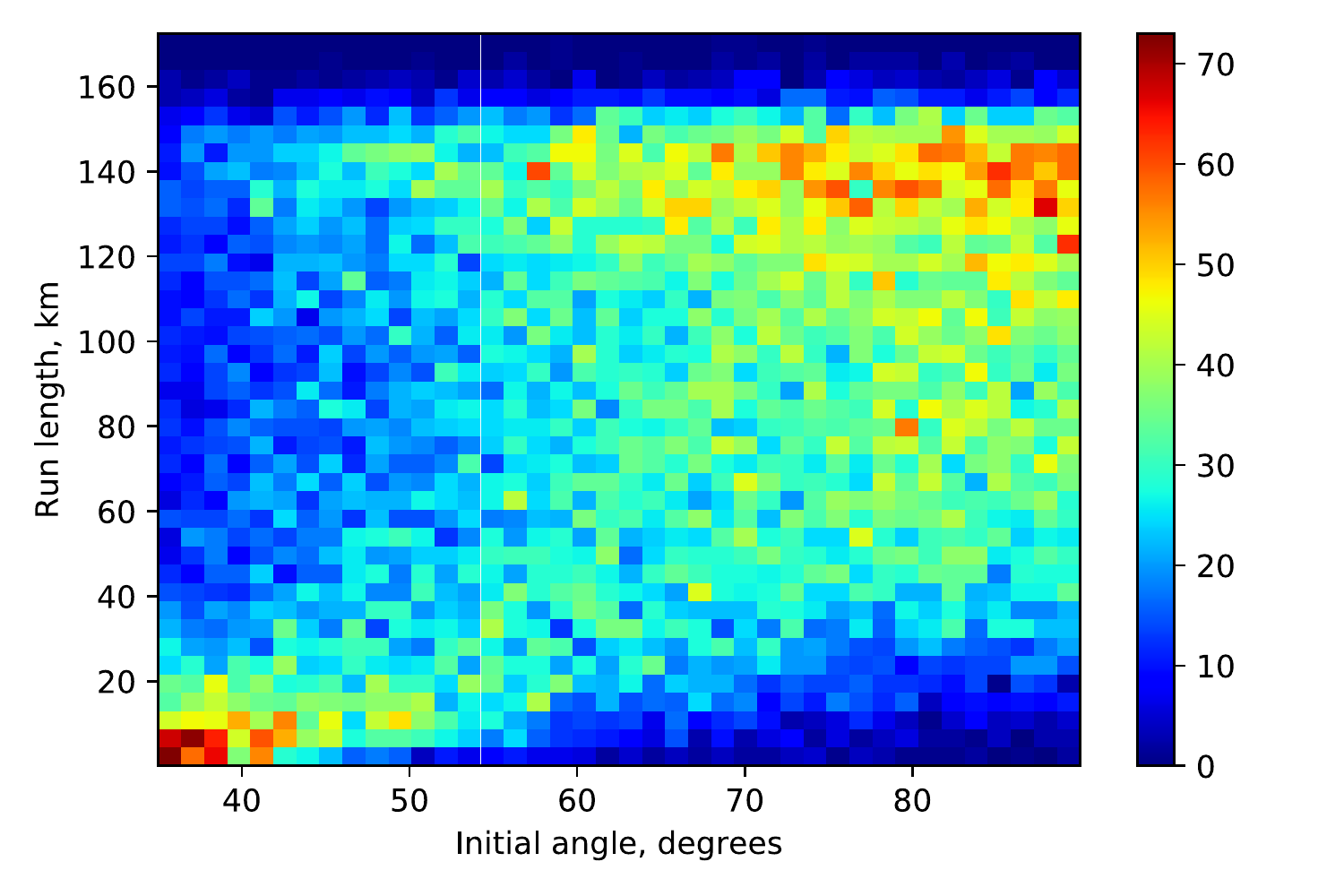}
        \caption{Full electron run length versus initial angle relative to field axis for ACCEPTED electrons. The initial energy was 20 keV. }
        \label{fig:length-vs-angle}
    \end{minipage}
\end{figure}

The resulting flat spectrum is quite peculiar. We see that the mono-energetic initial spectrum is transformed into an almost uniform "white noise" spectrum after evaporation. The result, while being a bit counter-intuitive, is reproduced quite well in different simulations. On the level of common sense, one can understand this spectrum in the following way: the escape time is proportional to the difference between the initial angle and the escape angle. The energy loss is proportional to escape time, which means it is also proportional to the angle difference. It means that the loss will be proportional to the initial angle. 

One must note that simulation does not directly track the time. Instead, we compute the total run length, which monotonously depends on time. Figures \ref{fig:loss-vs-length} and \ref{fig:length-vs-angle} show the dependence of energy loss on total flight path (which is a measure of time in this simulation) and dependence of flight path on the initial angle. The dependency is clear in Fig.~\ref{fig:loss-vs-length}. The dependency of length on the angle (Fig.~\ref{fig:length-vs-angle}) is fuzzier due to possible large-angle scatterings in quasi-elastic interactions, but the tendency still could be seen.

\begin{figure}
    \begin{minipage}{0.45\linewidth}
        \centering
        \includegraphics[width = \linewidth]{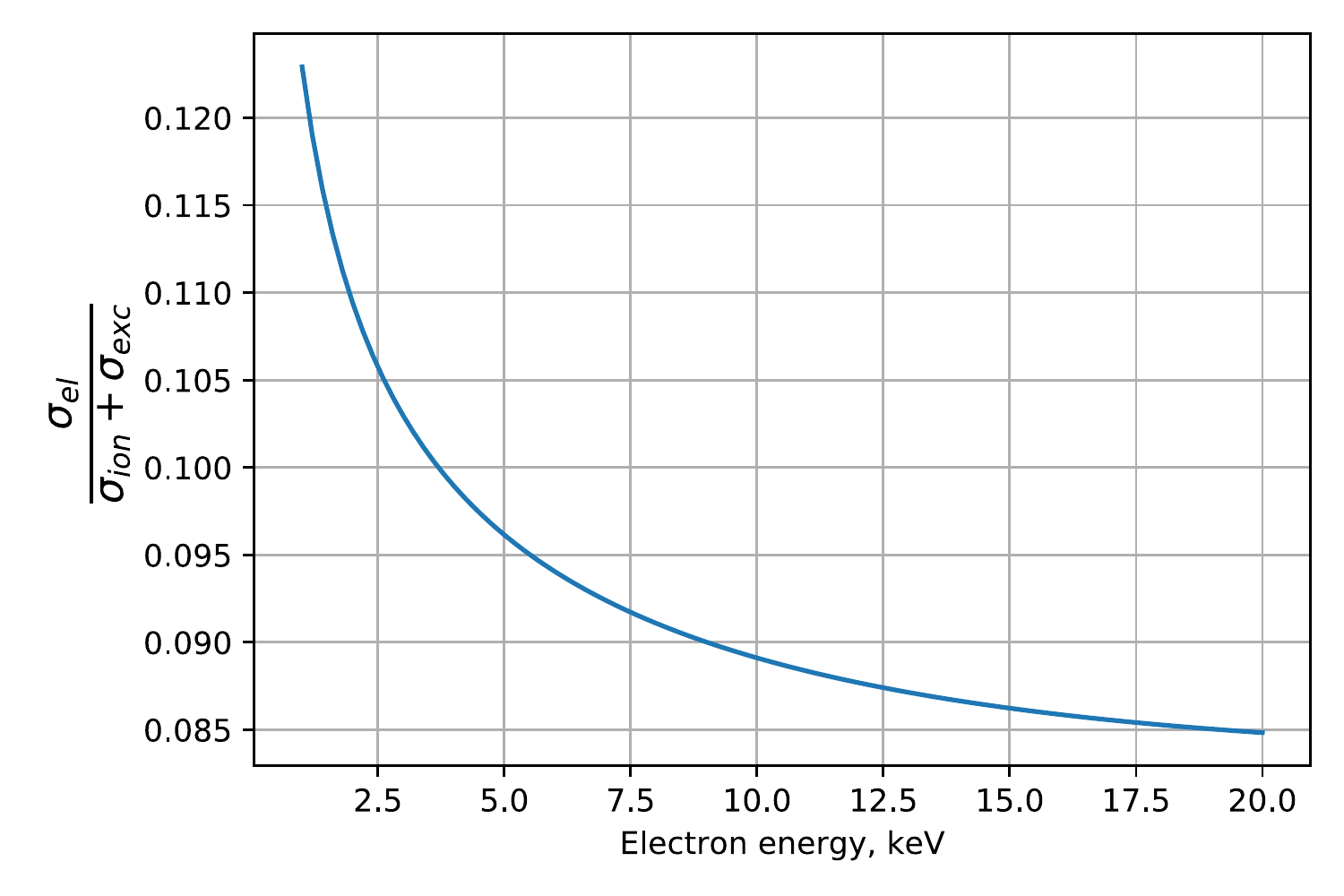}
        \caption{The ratio between quasi-elastic and excitation/ionization cross-sections.}
        \label{fig:cross-section-div}
    \end{minipage}
    ~
    \begin{minipage}{0.45\linewidth}
        \centering
        \includegraphics[width = \linewidth]{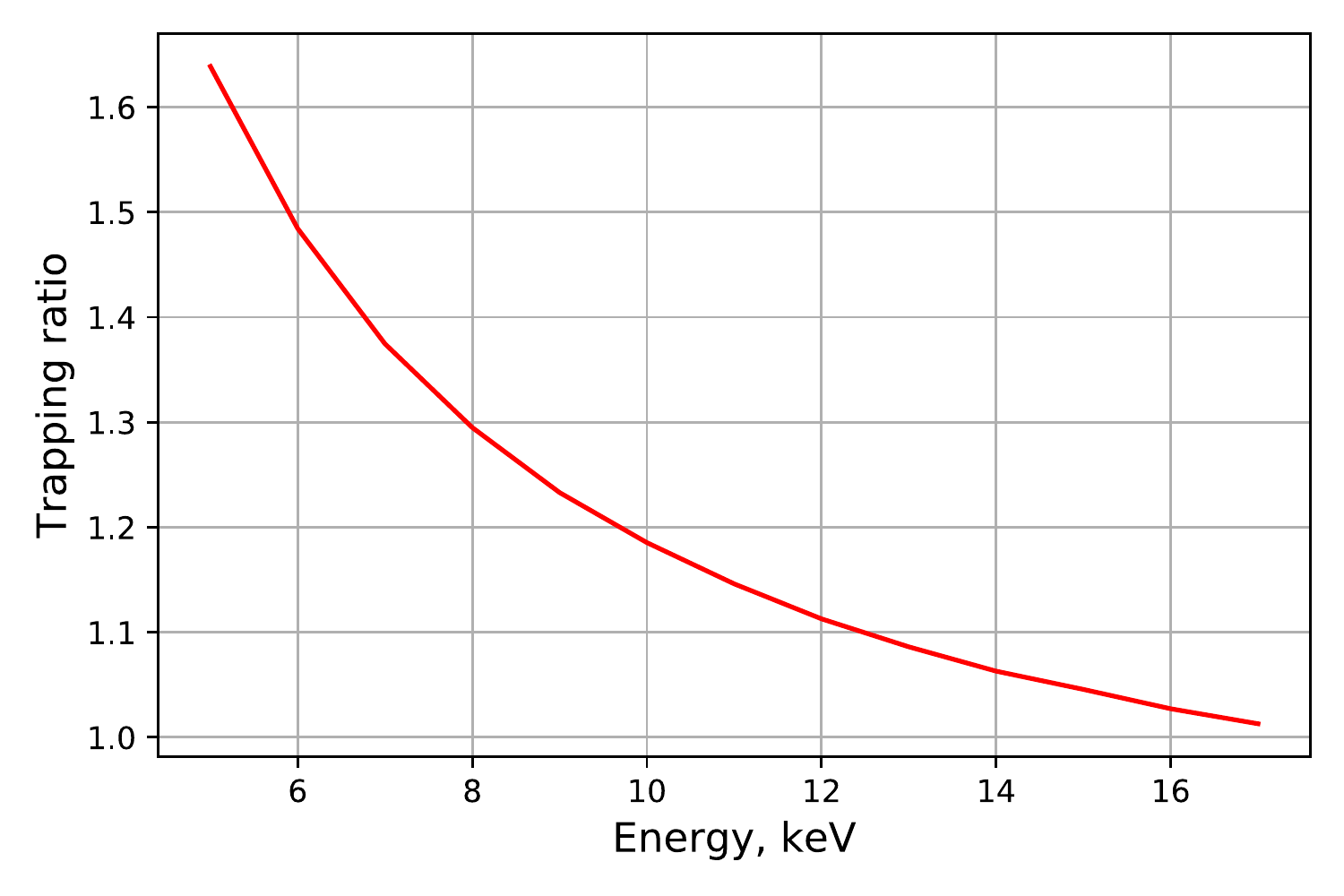}
        \caption{The ratio between tritium beta-spectrum with trapping and tritium original spectrum without trapping.}
        \label{fig:trap-ratio}
    \end{minipage}
\end{figure}

The simulation results show only small dependence on magnetic field parameters and no dependence on the pressure. The only effective parameter is the ratio between quasi-elastic and non-elastic cross-sections shown at Fig.~\ref{fig:cross-section-div} (some details on cross-section models could be found in \cite{Abdurashitov:2016nrv}). The ratio between cross-sections decreases slightly with energy, which explains the deviation from the constant spectrum at Fig.~\ref{fig:energy-spectra}. The ratio between tritium beta-spectrum with trapping and tritium beta-spectrum without trapping is shown at Fig.~\ref{fig:trap-ratio}. 

The precise experimental check of the trapping effect is not possible at the moment. The integral value of the effect is rather large (like it is shown at Fig.~\ref{fig:trap-ratio}), but the differential value in the measurement region is small and it is not possible to completely disentangle the trapping spectrum from the tritium beta-spectrum (because tritium beta-spectrum amplitude is not fixed). The only way to check trapping directly is to introduce a monochromatic electron radioactive source with isotropic angle distribution. There was an effort to do so with Kr source (see \cite{Belesev:2008zz} for details), but the power of the Kr source was not enough (relative to the residual tritium background) to draw any quantitative conclusions. The simulated spectrum with a full tracking (see Fig.~\ref{fig:track-compare}) was used in \cite{Aseev:2011dq} to search for electron neutrino and fast tracking was used in later articles in search for sterile neutrino. In all cases, there is a good agreement between the data and total model spectrum, which includes the trapping effect. This agreement could be used as an indirect experimental proof of simulations presented in this work.

Our study with different magnetic field configurations shows that the general shape of the spectrum remains the same independently of the magnetic trap shape. It means that the effect could be observed not only in Troitsk nu-mass experiment but in other cases as well.

One of such cases is the KATRIN experiment (\cite{Aker:2019uuj}). There is no significant magnetic trap in the KATRIN source due to the absence of the rear magnetic mirror, but at some scale, trapping is still possible in some places like pumping sections, where the pressure of tritium is still high enough to produce a significant amount of trapped electrons. The simulation code presented in this paper could be useful to calculate the spectrum of evaporated electrons there as well. 

Another possible case is the trapping of charged particles in Earth's magnetic field. Of course for this purpose the code should be changed, taking into account that trapped particles are mostly protons and not electrons. Also, it is possible, that one cannot discard transversal drift on such large scales.

\section{Conclusion}

The simulations presented in this article allow taking into account the effect of evaporation of electrons from a magnetic trap, which is noticeable for the Troitsk nu-mass experiment. The shape of the spectrum of electrons escaping the trap could be predicted with good precision. The simulations provide robust estimations of the spectrum. The Kotlin code allows computing millions of trapped particles per minute, which is by few orders of magnitude faster than standard solutions (like GEANT4). It allows us to get enough statistical precision for the final spectrum and to play around with the parameters. The same effect could be estimated in other cases of magnetic trapping as well.

The authors would like to thank people for consulting them on different stages of writing this article: Aino Skasyrskaya, Viktor Matushko, Polina Okuneva, and Oleg Komoltsev. Special thanks to Vladislav Pantuev and Peter Klimai both for valuable discussion and the manuscript proofreading. The JetBrains Research organization provided infrastructure and informational support for the authors.

This work is supported by the Ministry of Science and Higher Education of the Russian Federation under the contract 075-15-2020-778.

\bibliographystyle{unsrt}
\bibliography{references.bib}

\end{document}